\documentclass[12pt,a4paper,reqno]{article}

\usepackage{authblk}
\usepackage{geometry}
\usepackage[authoryear,round]{natbib}
\usepackage{amssymb}

\usepackage{graphicx}

\usepackage[utf8]{inputenc}
\usepackage[T1]{fontenc}
\usepackage{amsmath}

\newcommand{\E}{\mathbb{E}}
\newcommand{\R}{\mathbb{R}}
\newcommand{\M}{\mathbb{M}}
\newcommand{\dd}[1]{\text{d}#1}

\newcommand{\dv}{\dd{v}}
\newcommand{\suma}{\sum \limits_\alpha}
\newcommand{\va}{v_\alpha}
\newcommand{\wa}{w_\alpha}

\title{A structure-preserving particle discretisation for the Lenard-Bernstein collision operator}

\author[1,4]{S. Jeyakumar}
\author[2,3]{M. Kraus}
\author[1,5]{M. J. Hole}
\author[4]{D. Pfefferlé}

\affil[1]{\small Mathematical Sciences Institute, Australian National University, Acton ACT 2601, Australia}
\affil[2]{\small Max Planck Institute for Plasma Physics, Boltzmannstra{\ss}e 2, 85748 Garching, Germany}
\affil[3]{\small Technical University of Munich, Department of Mathematics, Boltzmannstra{\ss}e 3, 85748 Garching, Germany}
\affil[4]{\small The University of Western Australia, 35 Stirling Highway, Crawley WA 6009, Australia}
\affil[5]{\small Australian Nuclear Science and Technology Organisation, Locked Bag 2001, Kirrawee DC, NSW 2232, Australia}

\begin{document}

\maketitle

\begin{abstract}
Collisions are an important dissipation mechanism in plasmas.
When approximating collision operators numerically, it is important to preserve their mathematical structure in order to retain the laws of thermodynamics at the discrete level.
This is particularly challenging when considering particle methods.

A simple but commonly used collision operator is the Lenard-Bernstein operator, or its modified energy- and momentum-conserving counterpart. 
In this work, we present a macro-particle discretisation of this operator that is provably energy and momentum preserving. 
\end{abstract}

\tableofcontents

\pagebreak

\section{Introduction}\label{sec:introduction}

Structure-preserving numerical methods aim at preserving certain properties of a system of equations exactly at the discrete level.
Some examples for properties of interest are symmetries and conservation laws, Lagrangian or Hamiltonian structure, or compatibility with the laws of thermodynamics.
Preserving such structures is typically found to be advantageous for accuracy and robustness of numerical schemes, especially for strongly nonlinear problems and long-time simulations \citep{Hairer2006}.
This has also been recognised in plasma physics, and the last decade has seen vivid efforts towards the development of structure-preserving algorithms for problems such as magnetohydrodynamics, the Vlasov-Poisson and the Vlasov-Maxwell system (see e.g. \cite{Morrison:2017} and references therein).
So far, most work focused on dissipation-less systems, with dissipative systems, such as collisional kinetic systems, being considered only more recently.
However, dissipative effects, although often weak, are important for the correct simulation of physical behaviour over long simulation times.
Sometimes, the neglect of dissipative effects can cause numerical problems, e.g., when small structures emerge that cannot be resolved by the computational mesh.
In many cases, these structures are unphysical because dissipation would prevent their emergence in the first place.
Thus, the inclusion of dissipation is important not only for physical correctness but also because it can aid numerical robustness.

Work on the structure-preserving discretisation of Vlasov-like equations has mainly focused on particle-based methods.
In recent years, many authors worked on the ideal (non-dissipative) part of the problem, including \citet{Chen2011, Markidis2011, Squire2012, Evstatiev:2013, Qin2016, Burby2017, Kraus2017, Zhang2017, CamposPinto2022}.

After the discretisation of the ideal problem was well understood, focus shifted towards the structure-preserving discretisation of the collisional (dissipative) part.
While early work focused on grid-based methods (see e.g. \citet{Yoon2014, Taitano2015, Hirvijoki2017a, Kraus2017b, Shiroto2019}), structure-preserving discretisations for collision operators \textit{with particles} have been considered more recently.
\citet{hirvijoki2018metriplectic} consider an approach where the weights of the marker particles are varied, instead of their velocities.
\citet{Carrillo2020} and \citet{Hirvijoki2021} use finite-sized marker particles to discretise the Landau operator.
\citet{Mollen2021} and \citet{Pusztay2022} focus on projection/interpolation techniques for computing collision operators for particles. 
An alternative approach is that of \cite{Tyranowski2021}, which treats the collisions as a stochastic process, effectively modelling their underlying microscopic behaviour rather than the resultant macroscopic effects modelled by various collision operators.

The aim of this work is to provide a proof-of-concept for an alternative approach to structure-preserving particle methods for collisions, specifically for a conservative version of the operator of \citet{Lenard1958}. 
We employ the deterministic particle method (\cite{Chertock2017,Degond1990}) to obtain dynamical equations for the particle velocities. These are then regularised by evaluating the collisional flux on a smoothened representation of the distribution function that is obtained from a projection of the particles onto a spline basis. We show that this semi-discretisation maintains momentum and energy conservation exactly.
A midpoint discretisation in time is employed in order to retain these conservation properties also at the fully discrete level.

The structure of the paper is as follows. In Section~\ref{sec:cons_LB}, we detail the derivation of the conservative Lenard-Bernstein operator. In Section~\ref{sec:semi-discrete-operator}, the semi-discretisation of the operator is presented and its conservation properties are verified. Section~\ref{sec:time_discretisation} describes a possible time discretisation by the midpoint rule and proves that it retains the desired conservation properties. Section~\ref{sec:examples} shows several numerical tests and examples for the one-dimensional case, including some convergence results and verification of momentum and energy conservation. Finally, the paper is concluded with a discussion of current and future work.

\section{The conservative Lenard-Bernstein operator}
\label{sec:cons_LB}

The Lenard-Bernstein collision operator (\cite{Lenard1958}) is a scalar operator of the form
\begin{equation}\label{eq:LB_operator}
    C[f] = \nu\frac{\partial}{\partial v} \cdot \left(\frac{\partial f}{\partial v} + vf \right) ,
\end{equation}
where $f:\R^d \times [0,\infty) \to \R$ is the single-particle distribution function, $v \in \R^d$ is the velocity and $\nu$ is the collision frequency, which is assumed to be constant in time. In most applications, such a collision operator is coupled to the Vlasov-Poisson or Vlasov-Maxwell equations, and so the distribution function would also depend on the position variables. Here, however, we will ignore this dependency as we study the collision operator independently of the ideal dynamics and this operator acts purely in velocity-space.
Specifically, we solve the following differential equation:
\begin{equation}\label{eq:collisional-dynamics}
\partial_t f = C[f] .
\end{equation}
The Lenard-Bernstein operator is applicable in velocity dimensions $d = 1,2,3$, though collision operators which describe more physics effects, such as the Landau operator, may be preferred in two and three-dimensions in order to allow, for example, interchange of momentum between different components. The steady-state solution to~\eqref{eq:collisional-dynamics} for this operator is an $d$-dimensional Gaussian distribution. 

\subsection{Construction of the conservative operator}

The Lenard-Bernstein operator~\eqref{eq:LB_operator} preserves mass density, the zeroth moment of the distribution function. However, it does not preserve momentum density nor energy density, which are the first and second moments respectively, and whose conservation is crucial for obtaining physically correct results in numerical simulations. In order to enforce conservation of these quantities, we follow \citet{Kraus2013} (see also \cite{Filbet2003}) and modify the operator through an expansion as follows,
\begin{equation}\label{eq:cons_LB}
    C[f] = \nu\frac{\partial}{\partial v} \cdot \left(\frac{\partial f}{\partial v} + A_1f + A_2 vf \right) .
\end{equation}
In the following, we will see that the coefficients $A_m$ are functions of the moments of the distribution function $f$. In general, preserving $k$ moments of the distribution function will require an expansion including $k$ terms in the operator. 
The coefficients are then computed by requiring Equation~\eqref{eq:collisional-dynamics} to obey the respective conservation laws, which here are for momentum and energy.
Specifically, conservation of the $k$-th moment requires the following condition to hold: 
\begin{equation}
    \int v^k C[f] \dv = \nu \int v^k \left[ \frac{\partial}{\partial v} \cdot \left(\frac{\partial f}{\partial v} + A_1f + A_2 vf \right)\right] \dv = 0.
\end{equation}
Integrating this by parts, we obtain the following condition: 
\begin{equation}
    \nu \left[  v^k\left(\frac{\partial f}{\partial v} + A_1f + A_2 vf \right) \right]_{-\infty}^{+\infty} - k \nu \int v^{k-1}\left(\frac{\partial f}{\partial v} + A_1f + A_2 vf \right) \dv = 0 . \label{eq:cts_cons_derivation_1}
\end{equation}
Without loss of generality, we assume that $f$ and $\partial f/ \partial v$ approach zero as $|v| \to \infty$, so that the first term in Equation~\eqref{eq:cts_cons_derivation_1} is zero and we obtain the following condition:
\begin{equation}\label{eq:vanishing_collisional_flux}
    \int v^{k-1}\left(\frac{\partial f}{\partial v} + A_1f + A_2 vf \right) \dv = 0
\end{equation}
for $k = 1, 2. $ Integrating the first term by parts once again, this equation becomes:
\begin{equation}
    \int \left[ (k -1) v^{k-2} - A_1v^{k-1} - A_2 v^k \right] f \dv = 0,
\end{equation}
where the assumption that $f$ approaches zero as $|v|$ tends to infinity has been utilised once more.
Writing the moments as $M_m[f] = \int v^m f \dv$, we obtain the following conditions:
\begin{equation}
    (k - 1) M_{k-2} = A_1 M_{k-1} + A_2 M_{k}, \hspace{1em} k = 1, 2 .
\end{equation}
These conditions provide a linear system of equations that can be solved for the coefficients $A_1$, $A_2$:
\begin{align}
    \begin{split}
        A_1M_0 + A_2M_1 &= 0, \\
        A_1M_1 + A_2M_2 &= M_0. \label{eq:cts_cons_LB_system}
    \end{split}
\end{align}
The solution to the system of equations in \eqref{eq:cts_cons_LB_system} is: 
\begin{align}\label{eq:cts_A1_sol}
    A_1 &= \frac{M_0M_1}{M_0M_2 - M_1^2} = \frac{-u}{\varepsilon - u^2},\\
    A_2 &= \frac{-M_0^2}{M_0M_2 - M_1^2} = \frac{1}{\varepsilon - u^2},\label{eq:cts_A2_sol}
\end{align}
where $nu$ and $n\varepsilon$ are the momentum and energy density, respectively, and are related to the moments as follows:
\begin{equation}
    n = M_0 = \int f \dv, \hspace{1em}
    nu = M_1 = \int vf \dv, \hspace{1em}
    n\varepsilon = M_2 = \int v^2f \dv.
\end{equation}
Let us note that here, $n$, $u$, and $\varepsilon$ are just constants. However, in the general Vlasov-case, these quantities have a spatial dependency.
Upon inserting the expressions for $A_1, A_2$ into~\eqref{eq:cons_LB}, we obtain the following operator:
\begin{equation}
    C[f] = \nu\frac{\partial}{\partial v} \cdot \left(\frac{\partial f}{\partial v} + \frac{v-u}{\varepsilon - u^2} f \right), \label{eq:full_cons_LB}
\end{equation}
which can be seen as a conservative version of the Lenard-Bernstein operator~\eqref{eq:LB_operator}. This is the same operator as the one obtained by \cite{Filbet2003} and is closely related to the operator studied in \cite{Hakim2020}. 

\subsection{H-theorem}

We follow a similar strategy to \cite{Hakim2020} to demonstrate that the continuous operator satisfies a H-theorem. Let us denote the collisional flux by
\begin{equation}\label{eq:collisional-flux}
F[f]=\frac{\partial f}{\partial v} + A_1 f + A_2 vf ,
\end{equation}
so that 
\begin{equation}\label{eq:H-proof_eq1}
    \frac{\partial f}{\partial t} = C[f] = \nu \frac{\partial}{\partial v} \cdot F[f].
\end{equation}
The change in entropy is then
\begin{align}
    \nonumber
    \frac{dS}{dt} &= \frac{d}{dt} \int f \log f \dv
    = \int (1 + \log f ) \frac{\partial f }{\partial t} \dv
    = \nu \int (1 + \log f ) \frac{\partial}{\partial v} \cdot F \dv, \\
    &= - \nu \int \frac{1}{f} \frac{\partial f}{\partial v} \cdot F \dv
    = - \nu \int \frac{1}{f}|F|^2 \dv + \int (A_1 + A_2 v) F \dv ,
\end{align}
where integration by parts, the assumption that $F$ goes to zero as $|v| \to \infty$ and the substitution $\partial f / \partial v = F - A_1 f - A_2 v f$ were used.
By Equation~\eqref{eq:vanishing_collisional_flux}, the second term in the last expression is designed to vanish, namely
\begin{align*}
    \int (A_1 + A_2 v) F \dv &= 0.
\end{align*}
Hence, the entropy is monotonically dissipated (provided that $f$ is non-negative): 
\begin{equation}
    \frac{dS}{dt} = - \int \frac{1}{f} F^2 \dv \leq 0.
\end{equation}
The entropy is stationary when $F[f_{eq}]=0$, that is when $f_{eq}$ solves the PDE
\begin{align}
    \frac{\partial f_{eq}}{\partial v} = -\frac{v-u}{\varepsilon-u^2}f_{eq}.
\end{align}
The (unique) solution with conditions $f\overset{|v|\to \infty}{\longrightarrow}0$ and $\int f \dv=n$ is a $d$-dimensional shifted Gaussian distribution with mean $u$ and variance $\varepsilon-u^2$,
\begin{align}
    f_{eq}(v) = \frac{n}{(2\pi(\varepsilon-u^2))^{d/2}} \, e^{-\frac{1}{2}(v-u)^2/(\varepsilon-u^2)} .
\end{align}
Thus, the continuous operator of \eqref{eq:full_cons_LB} satisfies the H-theorem.

\section{Semi-discrete operator}
\label{sec:semi-discrete-operator}

In order to discretise the conservative Lenard-Bernstein operator in velocity-space, we need to introduce a second representation of the distribution function.
The particle-representation with Dirac delta distributions, which is usually used to solve the ideal part, is not differentiable and thus cannot be used to evaluate the collisional part.
Previous works regularised the collision operator by using finite sized particles (\cite{Carrillo2020, Hirvijoki2021}).
Here, we explore a different approach based on finite element or spline spaces of sufficient regularity.

The particle distribution function is given by
\begin{equation}\label{eq:klimontovich_dist}
f_p(v,t) = \suma \wa \delta(v - \va (t)),
\end{equation}
where $\{\va(t)\}_{\alpha=1}^N$ are the particle velocities which evolve over time, where $N$ is the number of particles. As the particle distribution function $f_p$ is non-differentiable, we use an $L^2$ projection of $f_p$ onto a set of differentiable basis functions $\{\varphi_j\}_{j=1}^M$ for $M \ll N$ as follows:
\begin{equation}\label{eq:projection_f}
f_s(v,t) = \sum_i \varphi_i(v) f_i(t)  = \sum_{i,j} \varphi_i(v) \, \M_{ij}^{-1} \suma \wa \varphi_j (\va(t)) ,
\end{equation}
where $\{f_i(t)\}$ are the coefficients of the projected distribution function, $f_s$, expressed in the basis $\{ \varphi_i\}$, and $\M_{ij} = \int \varphi_i \varphi_j \text{d}x$ are the elements of the corresponding mass matrix $\M$. The projected representation of the distribution function, $f_s(v)$, will be used as an auxiliary representation for the evaluation of the collision operator where differentiability is required. This type of projection also offers the benefits of smoothing the solution for appropriately-chosen basis functions $\{\varphi_j\}$. We note that Equation~\eqref{eq:klimontovich_dist} remains the primary representation of the distribution function in our method, and that Equation~\eqref{eq:projection_f} is only used in order to satisfy the differentiability requirements of the collision operator. 

To construct the semi-discretisation of the conservative Lenard-Bernstein operator, we return to its form in Equation~\eqref{eq:cons_LB}. We will discretise this equation first, and then derive the discrete coefficients $A_1$ and $A_2$ in terms of the discrete momentum and energy densities.

To discretise the conservative collisional dynamics,
\begin{equation}
    \frac{\partial f}{\partial t} = \nu\frac{\partial}{\partial v} \cdot \left(\frac{\partial f}{\partial v} + A_1 f + A_2 vf \right) ,
    \label{eq:cons_LB_PDE}
\end{equation}
we apply a deterministic particle method (see \citet{Chertock2017} for a review). Typically, deterministic particle methods are formulated for first order transport-type problems. They can, however, be adapted to the context of diffusion problems as shown by \citet{Degond1990}. Following this approach, we rewrite Equation~\eqref{eq:cons_LB_PDE}:
\begin{equation}
    \frac{\partial f}{\partial t} = \frac{\partial}{\partial v} \left(a(v, f) f  \right), \hspace{1em} a(v,f) = \nu \left( \frac{1}{f}\frac{\partial f}{\partial v} + A_1 + A_2 v \right) .
\end{equation}
This equation is approximately solved in terms of the particle distribution function~\eqref{eq:klimontovich_dist}, where the particle velocities $\{\va\}$ satisfy the following ordinary differential equations:
\begin{align}
    \dot{v}_\alpha = a(\va, f) = \nu \left(\frac{ 1}{f(\va)} \frac{\partial f}{\partial v}(\va) + A_1  + A_2 \va \right).
    \label{eq:particle_PDE}
\end{align} 

Let us note that the first term on the right-hand side is not well defined if the distribution function $f$ is replaced by its particle representation $f_p$. Therefore, we will use the projection shown in Equation~\eqref{eq:projection_f} instead and replace both instances of the distribution function $f$ with the projected distribution function $f_s$ to arrive at the following equation:
\begin{equation}\label{eq:velocity_ode}
    \dot{v}_\alpha = a(\va, f_s) = \nu \left( \frac{ 1}{f_s(\va)}\frac{\partial f_s}{\partial v}(\va) + A_1  + A_2 \va \right).
\end{equation}

The final step in obtaining the semi-discrete system of equations is to compute the coefficients $A_1$ and $A_2$. In analogy to the continuous case of Section~\ref{sec:cons_LB}, $A_1$ and $A_2$ are determined by requiring conservation of the discrete momentum and energy\footnote{Here, we have chosen to preserve the discrete moments in the particle-basis but it is also possible to derive a different scheme by imposing the conservation conditions on the projected moments.}:
\begin{align}
   \frac{d}{dt} \suma \wa \va &= \nu \suma \wa \left[  \frac{ 1}{f_s(\va)}\frac{\partial f_s}{\partial v}(\va) + A_1  + A_2 \va  \right]= 0, \label{eq:cons_1}\\
    \frac{1}{2}\frac{d}{dt} \suma \wa \va^2 &=  \nu \suma \wa \left[ \frac{ \va}{f_s(\va)} \frac{\partial f_s}{\partial v}(\va) + A_1 \va + A_2  \va^2 \right]= 0 .\label{eq:cons_2}
\end{align}
Upon introduction of the discrete mass, momentum, and energy densities,
\begin{align}
    n_h(\va) = \suma \wa, \hspace{1em}
    n_hu_h(\va) = \suma \wa \va , \hspace{1em}
    n_h\varepsilon_h(\va) = \suma \wa \va^2, 
\end{align}
we obtain a linear system of equations which can be solved to find the discrete $A_1$, $A_2$:
\begin{align}
    \begin{split}
        A_1n_h + A_2n_hu_h &= -\suma \wa \frac{f'_s(\va)}{f_s(\va)}, \\ 
        A_1n_h u_h + A_2n_h\varepsilon_h &=  -\suma \wa \va \frac{f'_s(\va)}{f_s(\va)}.
    \end{split}
    \label{eq:coefficient_lin_system}
\end{align}
The solution to this linear system is as follows:
\begin{align}\label{eq:discrete_A_solution}
    \begin{split}
        A_1 &= \frac{1}{n_h\varepsilon_h - n_hu_h^2} \suma \wa(u_h\va - \varepsilon_h)\frac{f'_s(\va)}{f_s(\va)}, \\
        A_2 &= \frac{1}{n_h\varepsilon_h - n_hu_h^2} \suma \wa (u_h - \va)\frac{f'_s(\va)}{f_s(\va)}.
    \end{split}
\end{align}
We note that these quantities implicitly depend on time through their dependence on the particle velocities $\{\va (t)\}$, and so must be recomputed at every timestep. 

We also note that in order for the projection to preserve the moments of the distribution function, it is sufficient that the functions $\{1, v, v^2\}$ are contained in $span \{\varphi_j\}$. This can be demonstrated by considering the example of conservation of mass, which requires: 
\begin{equation}
    \int 1 f_p \dv = \int 1 f_s \dv.
\end{equation}
To show this holds, we begin from the condition used to construct the projection: 
\begin{equation}\label{eq:proj_condition}
    \int \varphi_j f_p \dv = \int \varphi_j f_s \dv, \hspace{1em} \forall j \in 1, \dots, M.
\end{equation}
Let $1 \in span \{ \varphi_j \}$. Then, there exists a set of coefficients $\{c_j\}$ such that $1 = \sum_j c_j \varphi_j$. Multiplying both sides of~\eqref{eq:proj_condition} with the corresponding $c_j$ and summing over $j$, we have: 
\begin{equation}
    \sum_j c_j \int \varphi_j f_p \dv = \sum_j c_j \int \varphi_j f_s \dv, 
\end{equation}
which by linearity of the integral becomes:
\begin{equation}
     \int \sum_j c_j \varphi_j f_p \dv =  \int \sum_j c_j \varphi_j f_s \dv.
\end{equation}
Since we have that $1 = \sum_j c_j \varphi_j$, we then have that 
\begin{equation}
     \int 1 f_p \dv = \int 1 f_s \dv,
\end{equation}
which is the required condition for mass conservation. Momentum and energy conservation follow similarly from requiring the functions $v$ and $v^2$ to be in the span of basis $\{\varphi_j\}$. We note that the spline basis chosen in the numerical implementation, as detailed in Section 5, satisfies this property since it is a cubic basis. 

We remark that at this time, a proof of monotonic entropy dissipation for the semi-discrete Lenard-Bernstein operator remains elusive. We have, however, numerically demonstrated that our discretisation maintains this property in Section~\ref{sec:examples}.

\section{Time discretisation}
\label{sec:time_discretisation}

In this chapter, we describe the temporal discretisation of the system of ODEs~\eqref{eq:velocity_ode} by the implicit midpoint scheme, which is a possible choice for a method that maintains the discrete conservation laws exactly (up to machine-precision).
Let us start by introducing some notation.
Denote by $h$ a single timestep, by $t_n = t_0 + nh$ the time after the $n$-th timestep, and by $\va^n \approx \va (t_n)$ the corresponding particle velocity.
Further, let $\va^{n + 1/2} = (\va^n + \va^{n + 1})/2$ be the particle velocity at the midpoint and, following Equation~\eqref{eq:projection_f}, denote the spline representation of the distribution function at the midpoint by
\begin{align}
    f_s^{n+1/2}(v) = \sum_{i,j} \varphi_i (v) \M^{-1}_{ij}\suma \wa \varphi_j(\va^{n + 1/2}) .
\end{align}
With this, the implicit midpoint scheme for the system~\eqref{eq:velocity_ode} can be written as follows: 
\begin{equation}
    \va^{n+1} = \va^n + h a\left( \va^{n + 1/2}, f_s^{n+1/2} \right).
\end{equation}

In order to verify conservation of the total momentum, let us consider the particle momentum at the $n+1$-th timestep:
\begin{align}
    \suma \wa \va^{n+1}
    \nonumber
    &= \suma \wa \left[  \va^n + h a\left( \va^{n + 1/2}, f_s^{n+1/2} \right)\right]
    , \\
    \nonumber
    &= \suma \wa \va^n 
    + h \suma \wa a\left( \va^{n + 1/2}, f_s^{n+1/2} \right), \\
    &= \suma \wa \va^n + h \suma \wa \left[  \frac{ 1}{f_s^{n+1/2}}\frac{\partial f_s^{n+1/2}}{\partial v}\left(\va^{n + 1/2}\right) + A_1  + A_2 \va^{n + 1/2}  \right].
\end{align}
We observe that the sum in the second term of this equation is exactly given by the left-hand side of Equation~\eqref{eq:cons_1}, evaluated at $\va^{n+1/2}$, and so is equal to zero as \eqref{eq:cons_1} is satisfied for all times in the scheme by construction. Thus, the particle momentum is conserved exactly by the implicit midpoint scheme. 

A similar result can be demonstrated for the total particle energy. At the $n+1$-th timestep, we have: 
\begin{align}\label{eq:discrete-energy-conservation-1}
    \suma \wa \left( \va^{n+1} \right)^2
    \nonumber
    &= \suma \wa \left[ \va^n + h a \left( \va^{n + 1/2}, f_s^{n+1/2} \right) \right]^2, \\
    \nonumber
    &= \suma \wa (\va^{n})^2 + h^2 \suma \wa a\left( \va^{n + 1/2}, f_s^{n+1/2} \right)^2 \\
    & \hspace{2em} + 2h \suma \wa \va^n a\left( \va^{n + 1/2}, f_s^{n+1/2} \right).
\end{align}
Now, we split the last term and replace one $\va^n$ by using the implicit midpoint rule, i.e. $\va^n = \va^{n+1} - ha\left( \va^{n + 1/2}, f_s^{n+1/2} \right)$ to obtain the following:
\begin{align}\label{eq:discrete-energy-conservation-2}
\nonumber
2h & \suma \wa \va^n a\left( \va^{n + 1/2}, f_s^{n+1/2} \right) \\
\nonumber
&= h \suma \wa \va^n a\left( \va^{n + 1/2}, f_s^{n+1/2} \right) + h \nonumber
\suma \wa \va^n a\left( \va^{n + 1/2}, f_s^{n+1/2} \right) \\
\nonumber
&= h \suma \wa \va^n a\left( \va^{n + 1/2}, f_s^{n+1/2} \right) \\ 
\nonumber
& \hspace{2em} + h\suma \wa \left[ \va^{n+1} - ha\left( \va^{n + 1/2}, f_s^{n+1/2} \right) \right]a\left( \va^{n + 1/2}, f_s^{n+1/2} \right) \\
&= h\suma \wa \left(\va^n + \va^{n+1}\right) a\left( \va^{n + 1/2}, f_s^{n+1/2} \right) - h^2 \suma \wa a\left( \va^{n + 1/2}, f_s^{n+1/2} \right)^2.
\end{align}
The last term in this equation cancels the second term on the right-hand side of \eqref{eq:discrete-energy-conservation-1}, and we are left with:
\begin{align}\label{eq:discrete-energy-conservation-3}
    \suma \wa \left( \va^{n+1} \right)^2 &= \suma \wa \left(\va^{n}\right)^2 + 2h\suma \wa \va^{n + 1/2}a\left( \va^{n + 1/2}, f_s^{n+1/2} \right) .
\end{align}
The second term on the right-hand side of~\eqref{eq:discrete-energy-conservation-3} equals the expression in~\eqref{eq:cons_2} evaluated at the midpoint $\va^{n + 1/2}$, and therefore is zero. Thus, the particle energy is preserved exactly in time.

\section{Numerical results}
\label{sec:examples}

In this chapter, we present numerical experiments for the one-dimensional conservative Lenard-Bernstein operator using B-spline basis functions of arbitrary order for projecting the particle distribution function.
The implementation is based on the Julia programming language \citep{Julia-2017}, and the package is available publicly \citep{michael_kraus_2023_8123649}. 

\subsection{Convergence of the semi-discrete operator}

\begin{figure}
    \centering
    \includegraphics[width = 0.7\textwidth]{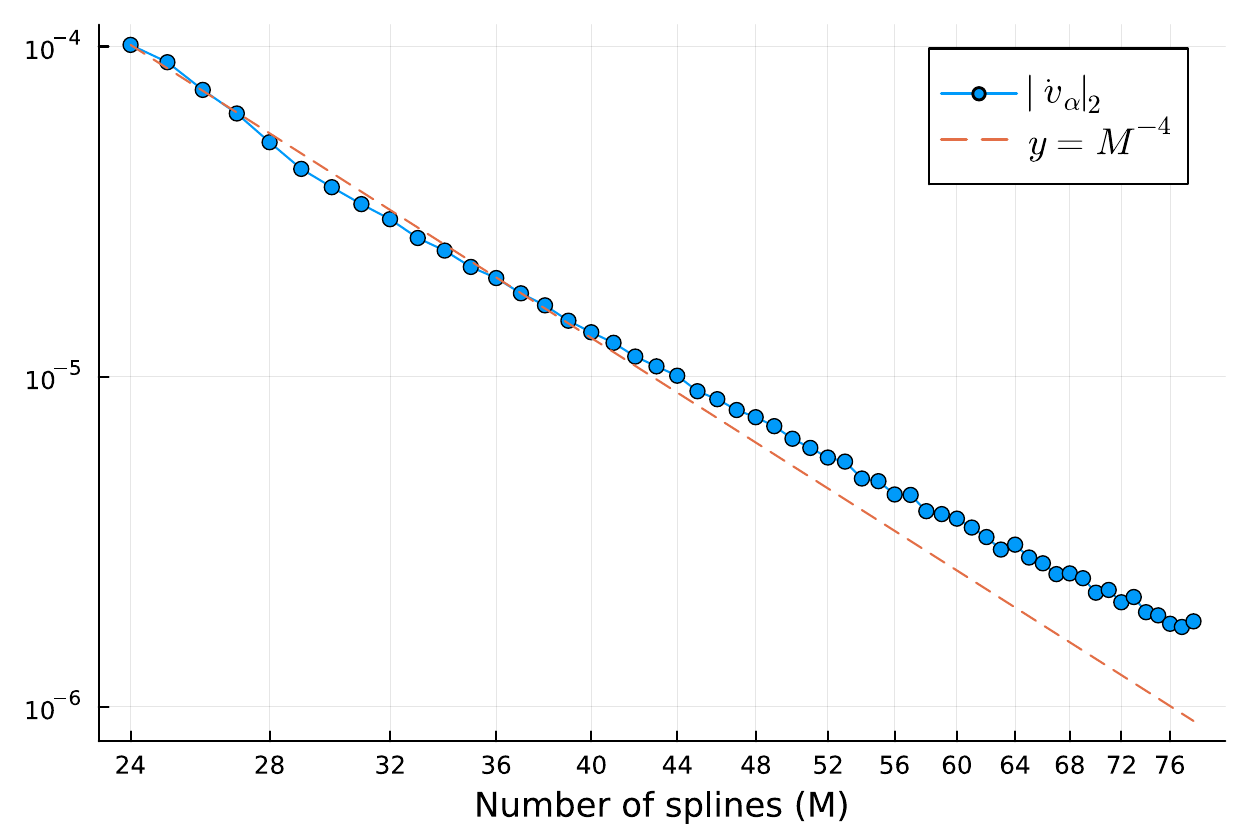}
    \caption{Convergence of the $L^2$ norm of the particle velocity gradient, $\| \dot{v}_\alpha \|_2$, when computed using a true Maxwellian $f$, against the number of splines. The dashed line shows the reference curve $y = x^{-4}$. }
    \label{fig:spline-convergence}
\end{figure}

We demonstrate convergence properties of the semi-discretisation under projection onto a B-spline basis and particle sampling. First, we verify that the semi-discretisation indeed preserves the Maxwellian as an equilibrium solution under projection. To this end, we compute the projection of an exact Maxwellian onto the spline basis as follows:
\begin{align}
    \int \sum_j f_j \varphi_j(v) \varphi_i(v) \dv = \int f_M(v) \varphi_i(v) \dv ,
\end{align}
where ${f_j}$ are the coefficients of the spline for which we are solving, and $f_M(v) = 1/\sqrt{2\pi} \exp(-v^2/2)$ is the Maxwellian of mean $\mu = 0$ and variance $\sigma^2 = 1$. Rearranging this expression, we obtain the following spline coefficients for the projected Maxwellian:
\begin{equation}\label{eq:numerical-example-1-coefficients}
    f_j = \M_{ij}^{-1} \int f_M(v) \varphi_i(v) \dv,
\end{equation}
where $\M_{ij} = \int \varphi_i \varphi_j \dv$ are the elements of the mass matrix $\M$ as before. The spline-projected representation of the Maxwellian distribution $f_M(v)$ is then given by $f_{s,M}(v) = \sum_j f_j \varphi_j(v)$ with the coefficients $f_j$ from~\eqref{eq:numerical-example-1-coefficients}. We use this projected Maxwellian to compute the right-hand side of Equation~\eqref{eq:velocity_ode}, as follows
\begin{equation}\label{eq:velocity_maxwellian_ode}
        \dot{v}_\alpha = \nu \left( \frac{ 1}{f_{s,M}(\va)}\frac{\partial f_{s,M}}{\partial v}(\va) + A_1  + A_2  \va \right),
\end{equation}
where the coefficients $A_1$ and $A_2$ are computed using the projected Maxwellian distribution $f_{s,M}$ in \eqref{eq:discrete_A_solution}. The $L^2$ norm of the time derivative of the particle velocities, $\| \dot{v}_\alpha \|_2$, can then be used to check if the Maxwellian is an equilibrium solution under projection, as this norm should approach zero with increasing spline resolution in such case. We compute this quantity for a range of spline resolutions, using a sample of $N = 100,000$ particles from a normal distribution where the sample is strictly used for evaluation of Equation~\eqref{eq:velocity_maxwellian_ode} and never for projection. Figure~\ref{fig:spline-convergence} shows the convergence of $\| \dot{v}_\alpha \|_2$ with an increasing number of splines, computed using cubic spline basis functions. As expected, the norm of the time derivative approaches zero with an increasing number of splines at a rate which corresponds to the order of the splines used (cubic splines have order $k = 4$). 

\begin{figure}
    \centering
    \includegraphics[width = 0.7\textwidth]{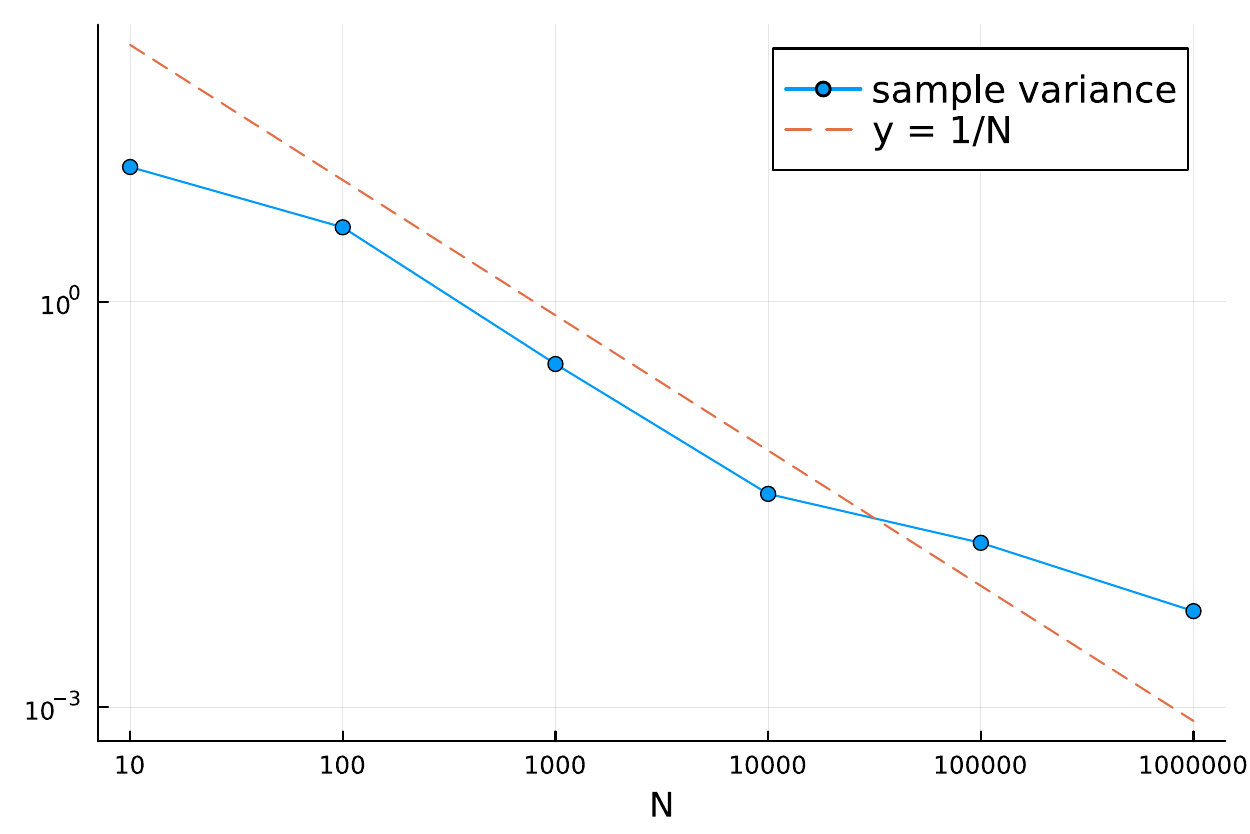}
    \caption{Convergence of the sample variance of the particle velocity time derivative, $\sum \dot{v}_\alpha ^2/N$, with the number of particles, shown on a logarithmic scale. The dashed line represents the reference curve $y = 1/N$.}
    \label{fig:particle-convergence}
\end{figure}

Secondly, we demonstrate convergence of the semi-discretisation under particle sampling. Here, we instead compute the sample variance of the particle velocity time derivatives, i.e. $\sum \dot{v}_\alpha ^2/N$, keeping the spline resolution fixed and varying the number of particles in the sample. Here, we directly project the particles to compute the spline representation of $f$, as per Equations~\eqref{eq:projection_f} and~\eqref{eq:velocity_ode}. The results of this are shown in Figure~\ref{fig:particle-convergence}, and we observe that the sample variance converges at a rate slightly above $1/N$, which is the expected convergence rate for the sample variance generated by statistical sampling methods. 

These two results demonstrate that the Maxwellian distribution remains an equilibrium solution under semi-discretisation, and the method demonstrates the expected convergence properties with both number of splines and particles.

\subsection{Relaxation of a shifted normal distribution}
\label{sec:first_example}

In the first example, we initialise the distribution function with a standard normal distribution shifted to the right by $\mu = 2$, obtaining a distribution with mean $\mu = 2$ and variance $\sigma^2 = 1$ as shown in the following equation: 
\begin{equation}
    f(v,t=0) = \frac{1}{\sqrt{2\pi}}e^{- \frac{1}{2}(v - 2)^2}.
\end{equation}
The particles are then initialised by independently and identically (iid) sampling from this distribution, for $N = 1000$ particles. The spline distribution is initialised by $L^2$ projecting the initial particle distribution onto the spline basis, with the spline coefficients computed as per Equation~\eqref{eq:projection_f}. A cubic-spline basis of 41 elements was used, with equally spaced knots on the velocity domain $v \in [-10, 10]$. The time integration is performed using the implicit midpoint scheme, which is of second order. A time-step of $h = 8 \times 10^{-4}$ and a collision frequency of $\nu = 1$ was used. The simulation was run until a final normalised time of $t = 1$, corresponding to $1250$ timesteps. 

\begin{figure}
    \centering
\includegraphics[width=\textwidth]{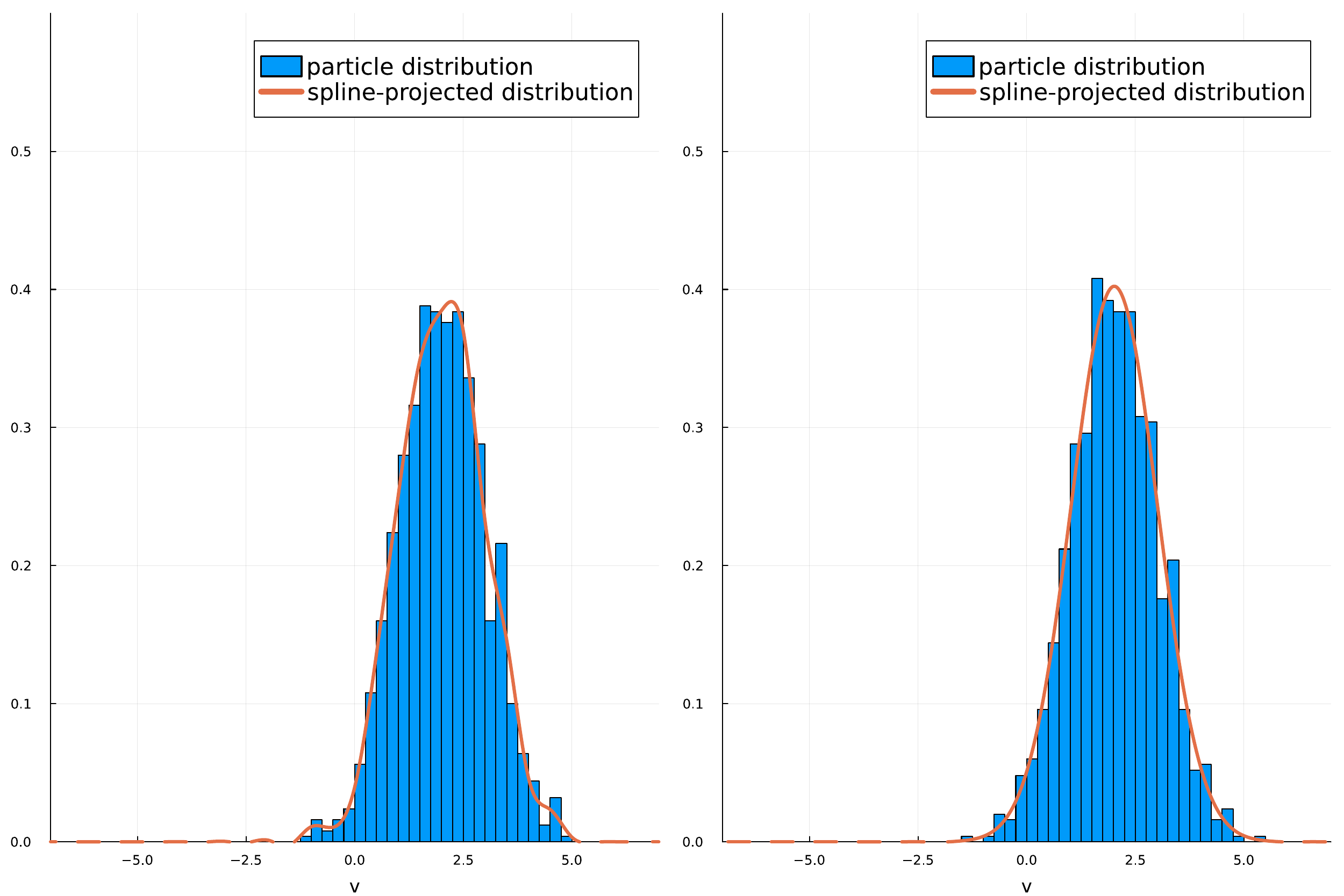}
    \caption{The initial and final distributions when the initial condition is chosen to be a normal distribution of mean $\mu = 2$ and variance $\sigma^2 = 1$, for a sample of $N=1000$ particles.} 
    \label{fig:shifted_normal_example}
\end{figure}

\begin{figure}
    \centering
    \begin{minipage}{.45\textwidth}
    \includegraphics[width=\textwidth]{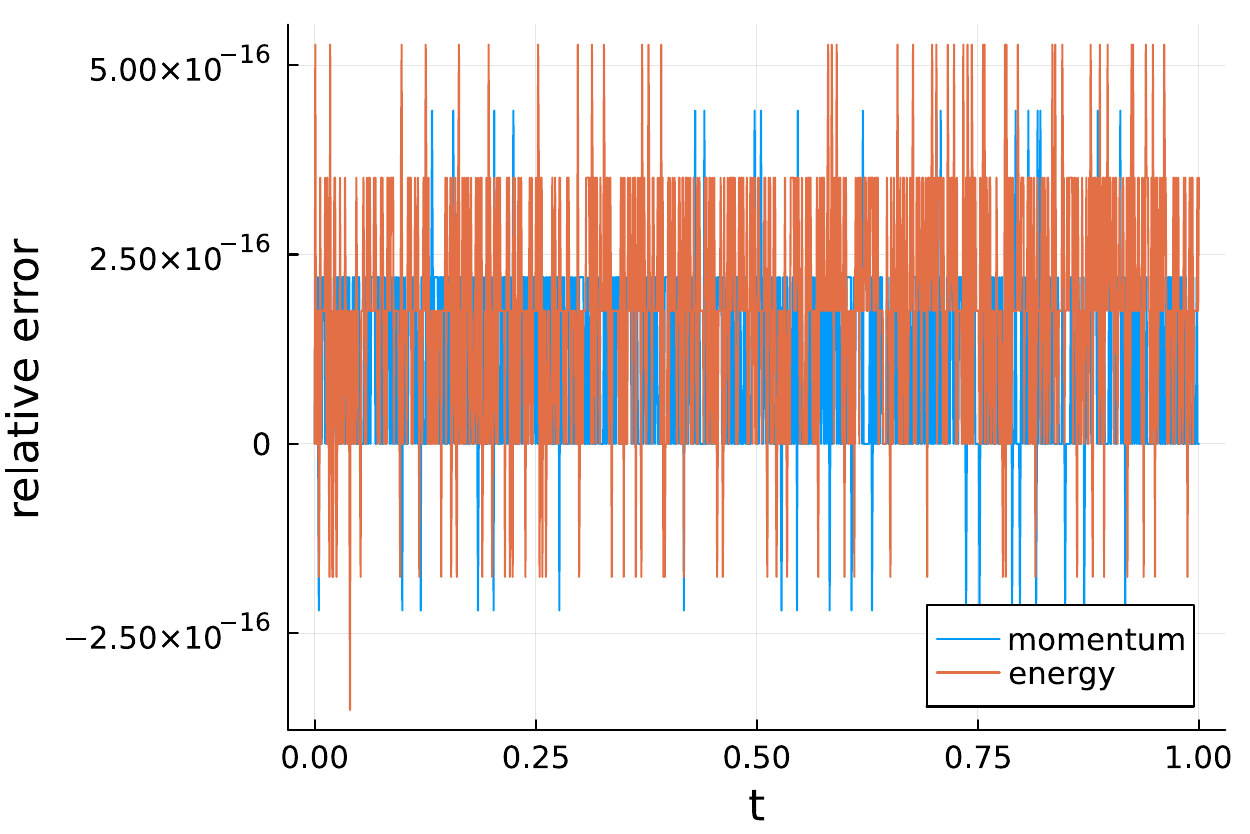}
    \caption{Energy and momentum conservation during the simulation, for the initial condition of a shifted normal distribution}
    \label{fig:conservation}
    \end{minipage}
    \begin{minipage}{.06\textwidth}
    \hfill
    \end{minipage}
    \begin{minipage}{.45\textwidth}
    \includegraphics[width=\textwidth]{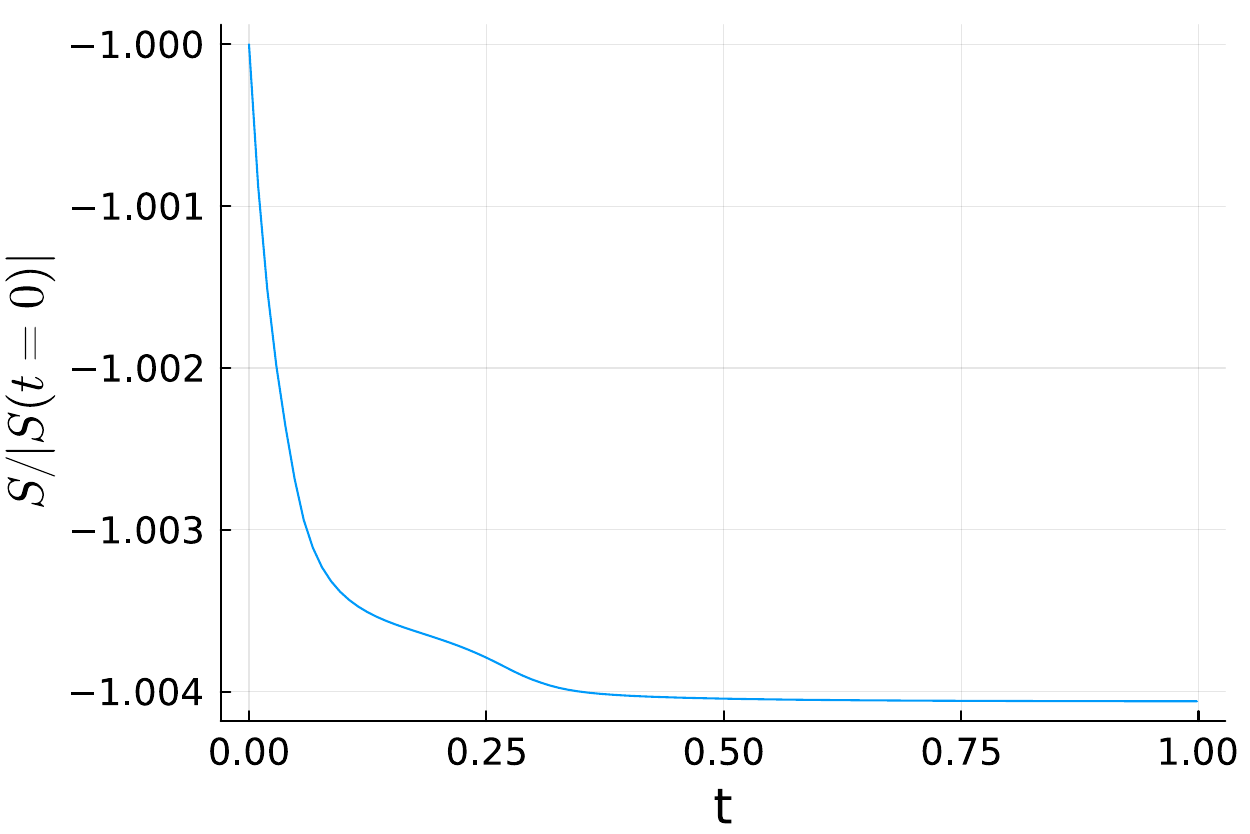}
    \caption{Evolution of the normalised entropy, i.e. $S/|S(t=0)|$ where $S = \int f_s \log{f_s} dv$, during the simulation}
    \label{fig:entropy_ex1}
    \end{minipage}
\end{figure}

In the initial condition, shown on the left of Figure~\ref{fig:shifted_normal_example}, we observe slight variations in the spline-projected distribution function due to the sampling error of the particles. In the final distribution, we observe the expected behaviour of the projected distribution approaching an exact normal with the initial variations smoothed out, and due to the momentum conservation, the mean of the distribution stays at the same value ($v = 2$). Here, the particle momentum and the energy are conserved up to machine precision. The evolution of the energy and momentum error are shown in Figure~\ref{fig:conservation}. The evolution of the entropy, $S = \int f_s \log{f_s} dv$, is shown in Figure~\ref{fig:entropy_ex1} (normalised by the magnitude of its initial value), and we observe the monotonic dissipation of the entropy over the course of the simulation as expected. 

\begin{figure}
    \centering
    \includegraphics[width=\textwidth]{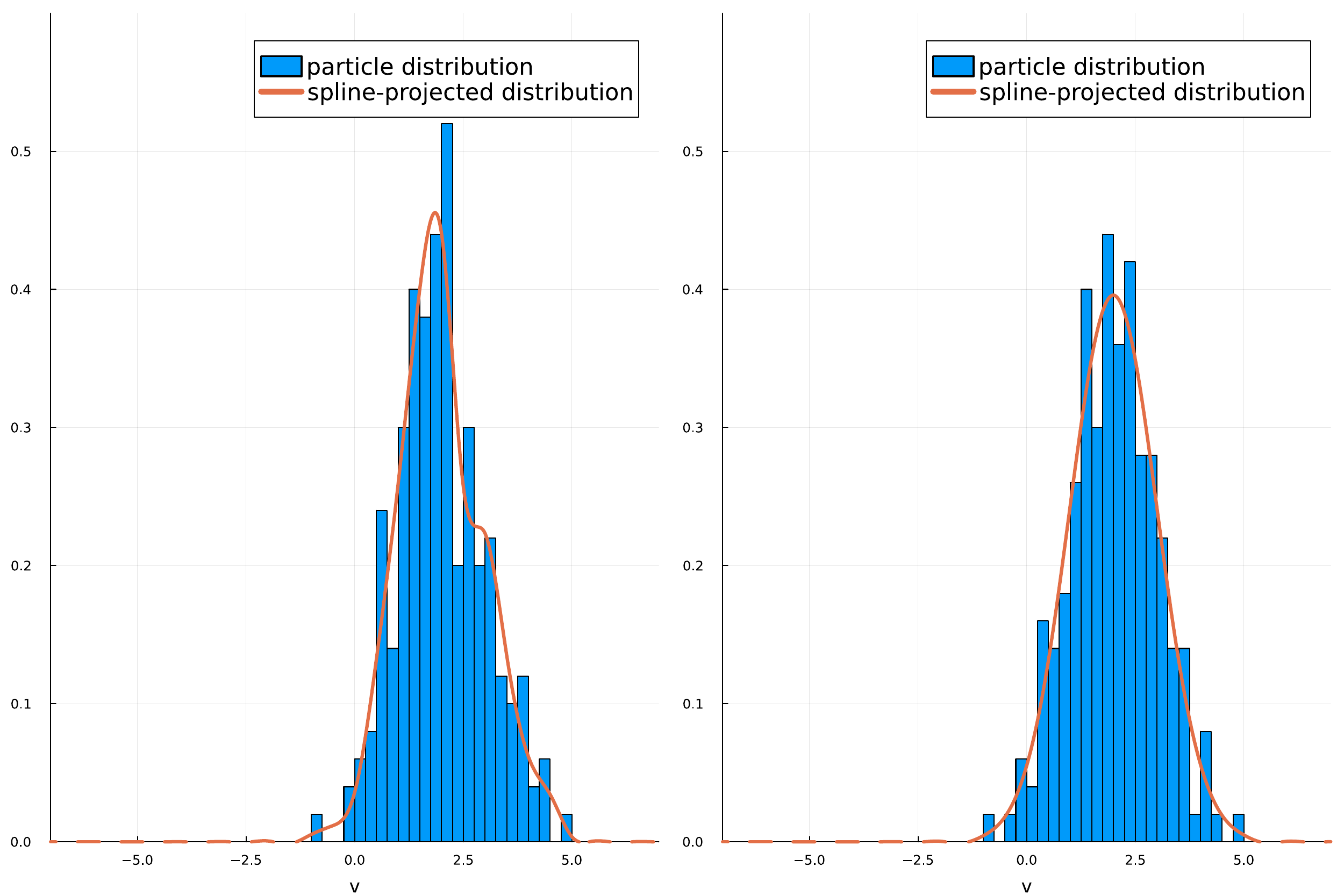}
    \caption{The initial and final distributions when the initial condition is chosen to be a normal distribution of mean $\mu = 2$ and variance $\sigma^2 = 1$, for a sample of $N = 200$ particles.}
    \label{fig:shifted_normal_ex2}
\end{figure}

We also observe that the method works well even for small numbers of particles. Results for the same simulation using a sample of $N=200$ particles instead are shown in Figure~\ref{fig:shifted_normal_ex2}. The particle energy and momentum are again conserved up to machine precision in relative error, with similar behaviour as shown in Figure~\ref{fig:conservation}.

\subsection{Relaxation of a bi-Maxwellian distribution}

\begin{figure}
    \centering
    \includegraphics[width=\textwidth]{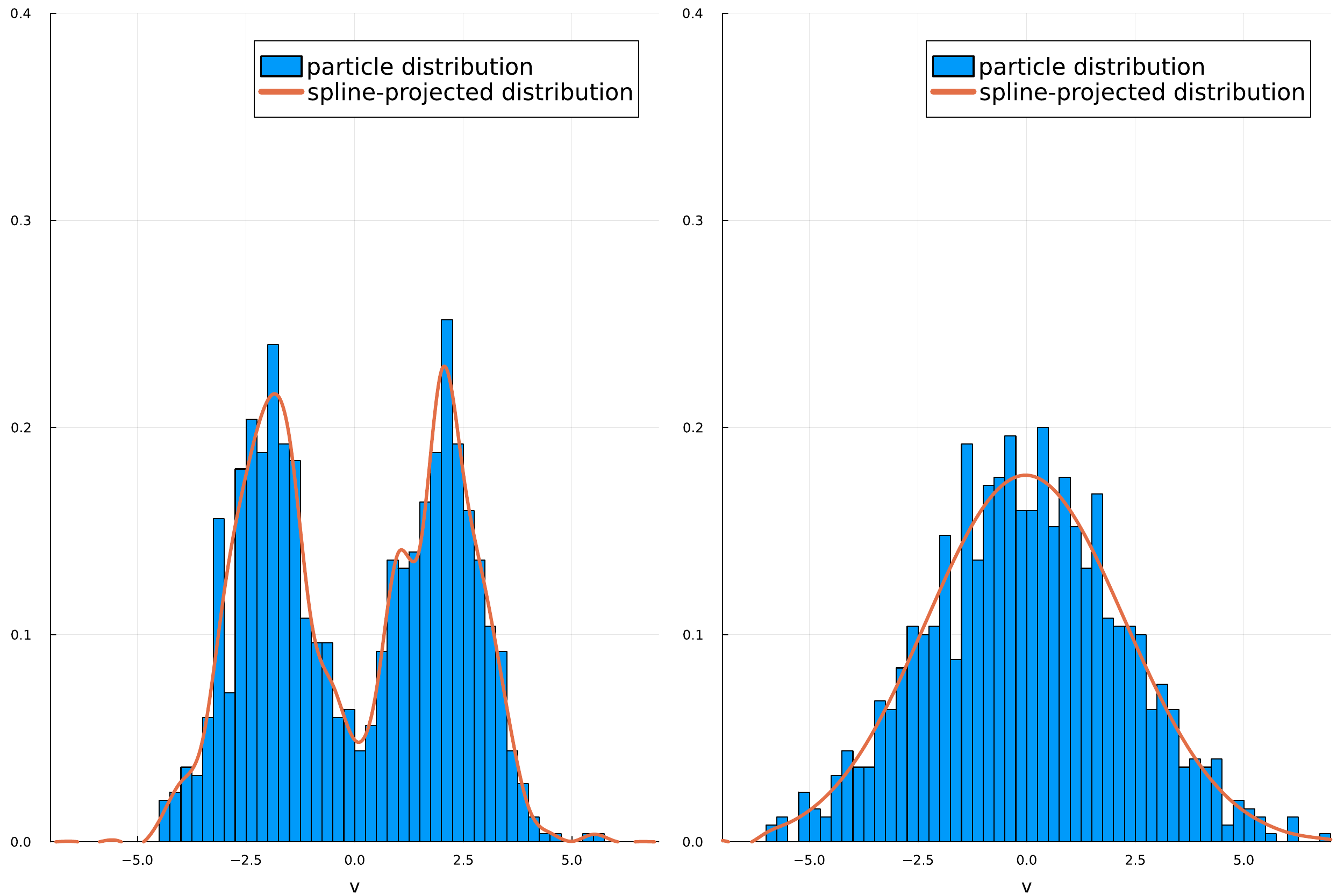}
    \caption{The initial and final distribution functions when the initial condition is chosen to be a bi-Maxwellian, in both particle and spline bases, for $N=1000$ particles.}
    \label{fig:double_maxwellian_example}
\end{figure}

\begin{figure}
    \centering
    \includegraphics[width=0.7\textwidth]{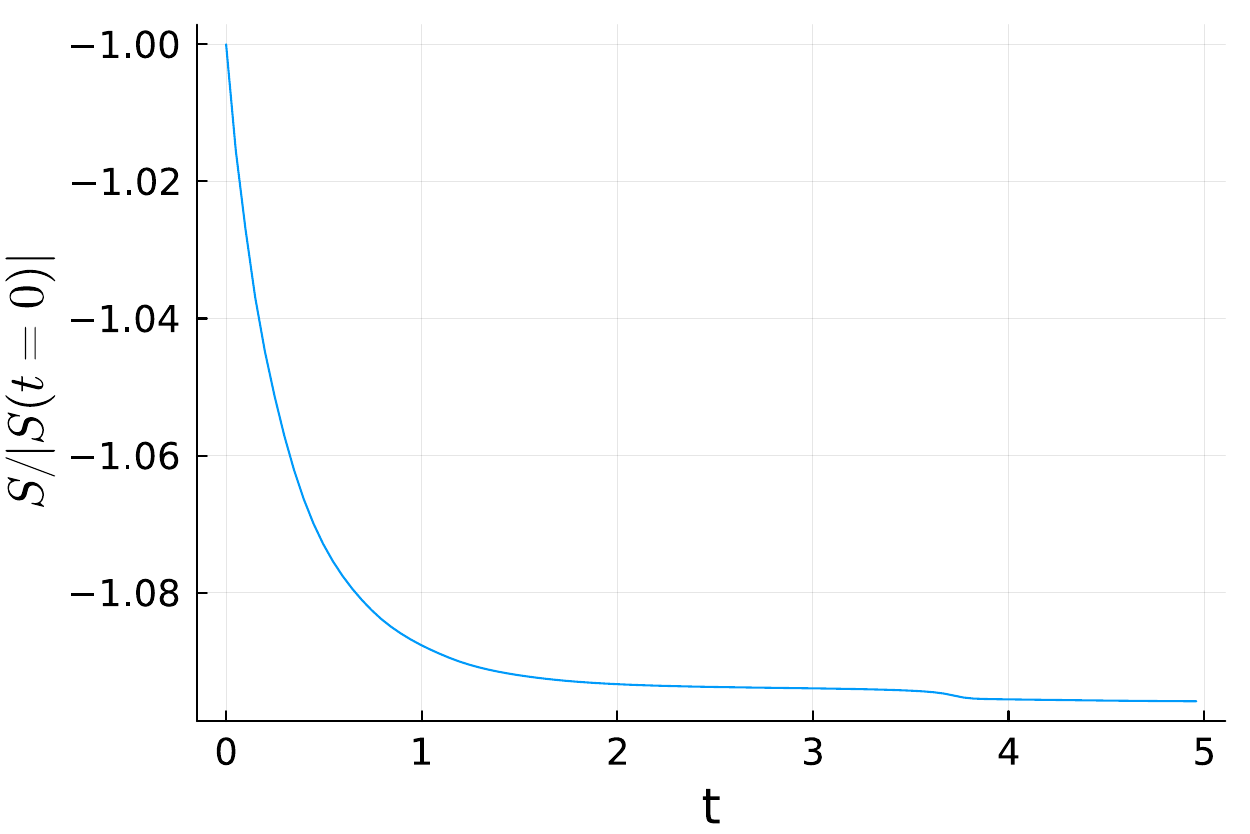}
    \caption{Evolution of the normalised entropy over the simulation, where the initial condition is a bi-Maxwellian.}
    \label{fig:entropy_ex2}
\end{figure}

Next, we consider a double Maxwellian distribution as the initial condition. Each peak is a standard normal distribution which has been shifted from the origin by $v = \pm 2$ as per the following equation: 
\begin{equation}
    f(v,t=0) = \frac{1}{\sqrt{2\pi}} \left( e^{- \frac{1}{2}(v - 2)^2} + e^{- \frac{1}{2}(v + 2)^2}\right),
\end{equation}
and the sampled distribution is shown in Figure~\ref{fig:double_maxwellian_example} on the left. The sample for the particle distribution function is again of size $N = 1000$ particles. We use the identical numerical setup as the previous example, in particular the same time-step of $h = 8 \times 10^{-4}$ and the same collision frequency of $\nu = 1$. The final distribution obtained after time integration until $t = 5$ is shown in Figure~\ref{fig:double_maxwellian_example} on the right. Again, the equilibrated distribution is a Gaussian with equal mean and variance to the initial condition. The particle energy is conserved up to machine precision in relative error. The particle momentum is conserved up to a relative error on the order of $10^{-14}$. The behaviour of the two quantities over time is oscillatory and similar to Figure~\ref{fig:conservation}. The normalised entropy also decreases monotonically in time, and this is illustrated in Figure~\ref{fig:entropy_ex2}.


\subsection{Relaxation of a uniform distribution}

\begin{figure}
    \centering
    \includegraphics[width=\textwidth]{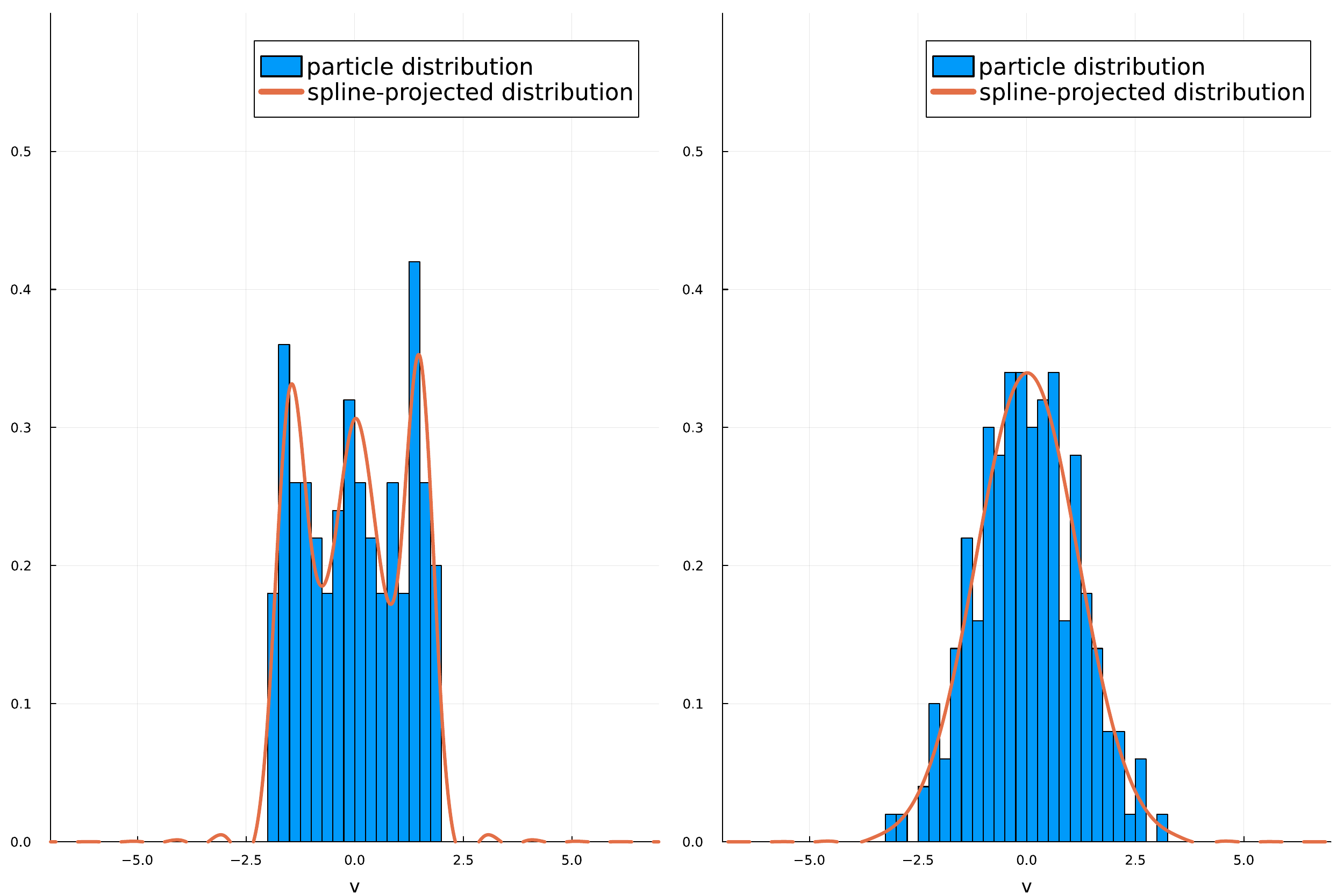}
    \caption{The initial and final distribution functions for the case of a uniform initial condition, with the initial condition shown on the left and the final result shown on the right. A sample of $N=200$ particles is used. }
    \label{fig:initial_final_dists_uniform}
\end{figure}

\begin{figure}
    \centering
    \includegraphics[width=0.7\textwidth]{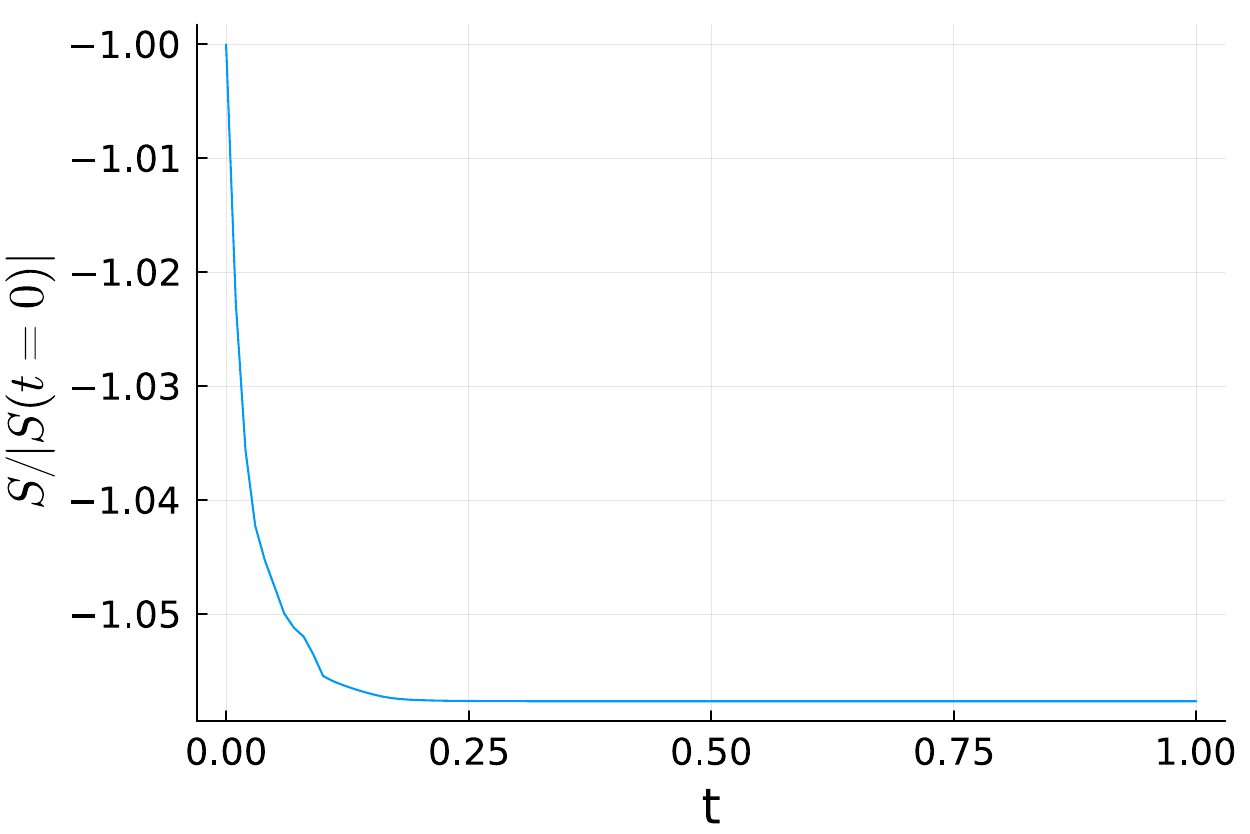}
    \caption{Evolution of the normalised entropy over the simulation, where the initial condition is a uniform distribution.}
    \label{fig:entropy_ex3}
\end{figure}

In the last example, we initialise the distribution function with a uniform distribution shifted and scaled to the interval $v \in [-2, 2]$, i.e.: 
\begin{equation}
    f(v, t=0) = \begin{cases}
        \frac{1}{4}, \text{ if } v \in [-2, 2],\\
        0, \text{ else.}
    \end{cases}
\end{equation}
A sample of $N = 200$ particles is taken from this distribution. A timestep of $h = 1 \times 10^{-4}$ is used and the final integration time is $t = 1$. All other parameters are kept the same.  Figure~\ref{fig:initial_final_dists_uniform} shows the initial and final distributions obtained in the simulation, demonstrating again the expected result that the initial distribution relaxes to a normal distribution. The resultant particle distribution function retains the same mean up to a relative error on the order of $10^{-14}$ (which is equivalent to momentum being preserved at this level). The energy is preserved to a relative error on the order of $10^{-15}$. The entropy decreases monotonically as expected and is illustrated in Figure~\ref{fig:entropy_ex3}. We note that the method performs as well here as it does in the other examples despite this being a more challenging case, as the uniform distribution is discontinuous and therefore not amenable to being represented using B-spline basis functions. 


\subsection{Remarks}

It is important to ensure that the chosen velocity domain for a simulation is sufficiently large such that no particle leaves this domain at any time. There is no sensible method for returning a particle to the simulation domain, as the true velocity space domain for this problem is infinite. Practically, a particle leaving the domain will return zero when the spline projected distribution is evaluated on the particle's velocity, which leads to the evaluation of \eqref{eq:velocity_ode} becoming undefined due to division by zero.

\section{Conclusion}

In this work, we have outlined the development of structure-preserving particle-based algorithms for the simulation of collision operators. While the approach itself is general, it has been specifically applied to a conservative version of the Lenard-Bernstein operator. We have derived an energy- and momentum-preserving particle discretisation for this operator, and the implicit midpoint method is shown to exactly preserve these quantities in time as well. We have demonstrated the convergence properties of the semi-discretisation under the projection used, as well as under particle sampling. Numerical examples for the one-dimensional case demonstrated the viability of the method and verified its conservation properties. The method is implemented in the Julia language with the respective repository available at \citep{michael_kraus_2023_8123649}. The proposed method can be coupled to any Vlasov-Poisson or Vlasov-Maxwell particle solver, and future research will detail such a coupling and its benefits. 

Currently, we are adapting the approach presented here for the Landau collision operator using the metriplectic formulation, and the results will be reported in a follow-up paper. 

\appendix 

\section{Time-evolution of cumulants}

One useful fact about the steady-state solution of the Lenard-Bernstein operator being a normal distribution is the fact that its cumulants of order three and above are all zero. The cumulant is a closely related quantity to a moment, being defined through a cumulant generating function which is obtained by taking the natural logarithm of the moment generating function of the distribution. The cumulant generating function for a normally-distributed random variable $X \sim N(\mu, \sigma^2)$ is given by
\begin{equation*}
    K(s) = \log \mathbb{E}\left[ \exp{sX} \right] = \mu s + \frac 1 2 \sigma^2 s^2,
\end{equation*}
where the cumulants are the coefficients of the Taylor expansion in $s$, $\kappa_n=K^{(n)}(0)$. In this instance, the cumulant generating function has no terms at order three and above, implying that the corresponding cumulants of the normal distribution are zero. We can also see that the first and second cumulants are simply the mean and variance, respectively\footnote{This is true in general for all probability distributions which have well-defined first and second moments, not only for the normal distribution.}. For ease of computation, the third and fourth cumulants can be related to central moments of the random variable (those centred around the mean) through the following relations:
\begin{align*}
    \kappa_3(X) &= \E \left[ (X - \E(X))^3 \right], \\
    \kappa_4(X) &= \E \left[ (X - \E(X))^4 \right] - 3\left(\E \left[ (X - \E(X))^2 \right]\right)^2,
\end{align*}
where $\kappa_3(X)$ and $\kappa_4(X)$ are the third and fourth cumulants, respectively. In the discrete setting, the behaviour of the discretised cumulants can act as a quantitative check of how close the solution is to the known solution. 

In fact, the time-evolution of the cumulants can be solved analytically in the case where the coefficients $A_k$ of the Lenard-Bernstein collision operator are held fixed. Let the moment-generating function be the Wick-rotation of the Fourier transform of the distribution:
\begin{align*}
    M(s;t) = \int e^{sv}f(v,t)dv = \E(e^{sX_t}) = e^{K(s;t)}.
\end{align*}
With this, moments of the distribution function are the Taylor coefficients of the moment-generating function $M_k(t)=\partial_s^{(k)}M(0;t)=\E(X_t^k)$.
If we write the Lenard-Bernstein collision operator as 
\begin{align*}
   C[f]=\nu\partial_v(\partial_vf + \sum_{k=0}^p A_k v^k f),
\end{align*}
then,  after integrating (and neglecting boundary terms originating from integration by parts), the relaxation equation $\partial_t f=C[f]$ becomes a PDE for $M(s;t)$, namely
\begin{align*}
    \nu^{-1}\partial_t M = s^2 M - s\sum\limits_{k=0}^p A_k \partial_s^{(k)}M ,
\end{align*}
where the substitution $\int e^{sv}v^k f dv=\partial_s^{(k)}M$ was exploited. 
By setting $s=0$ we see that the zeroth-moment is conserved,
\begin{align*}
    \frac{d}{dt} M_0 = 0 \iff M_0(t)=m_0 ,
\end{align*}
which can be attributed to the collision operator being a divergence.

The construction above is fairly general in the sense that the number of terms in the collision operator is capped by an arbitrary $p$. There are two distinct questions to address analytically.
\begin{enumerate}

    \item For arbitrary $p$, the steady-state moment-generating function $M_\infty(s)=\lim\limits_{t\to \infty}M(s;t)$ can be inferred from the ordinary differential equation (ODE)
    \begin{align*}
        \sum_{k=0}^p A_k M _{\infty}^{(k)} = s M_{\infty} .
    \end{align*}
    In particular if $p=1$, we have a first-order ODE 
    \begin{align*}
        M'_\infty = \frac{s-A_0}{A_1} M_\infty \iff K_\infty'=\frac{s-A_0}{A_1} ,
    \end{align*}
    together with the requirement that $K_\infty(0)=\ln 1 =0$. The solution is the cumulant-generating function of a Gaussian with mean $\mu=-A_0/A_1$ and variance $\sigma^2=1/A_1$:
    \begin{align*}
        K_\infty(s) =  \mu s + \frac{1}{2}\sigma ^2 s^2.
    \end{align*}

    \item In the case where $p=1$, normalising time to the collision frequency multiplied by the variance $t\mapsto t/(\sigma^2\nu)$, the relaxation equation in terms of the cumulant-generating function is
\begin{align*}
\partial_t K + s\partial_s K = \mu s + \sigma^2s^2 ,
\end{align*}
which is a linear non-homogeneous first-order PDE.  We apply the method of characteristics. Let the curve $s(\tau)$ and $t(\tau)$ and $\kappa(\tau)=K(s(\tau),t(\tau))$ be such that
\begin{align*}
    \frac{d t}{d\tau} &= 1, \\
    \frac{d s}{d\tau} &= s(\tau),\\
    \frac{d\kappa}{d\tau}  &= \frac{d}{d\tau} K(s(\tau),t(\tau)) 
    = \partial_s K \frac{ds}{d\tau} + \partial_t K\frac{dt}{d\tau}
    = \mu s(\tau) + \sigma^2 s(\tau)^2 ,
\end{align*}
with initial conditions $s(0)=s_0$, $t(0)=0$, and $\kappa(0)=K(s_0,0)=\mu s_0 + \frac{1}{2}\sigma^2 s_0^2+ R(s_0)$. The solutions are
\begin{align*}
    t(\tau)&=\tau,\\
    s(\tau;s_0)&=s_0 e^\tau\\
    \kappa(\tau;s_0) 
    &= \mu s_0 e^\tau + \frac{1}{2}\sigma^2 s_0^2 e^{2\tau} + R(s_0) .
\end{align*}
Inverting $s(\tau;s_0)$ and $t(\tau;s_0)$, we obtain the time-evolution of the cumulant-generating function as 
\begin{align*}
    \tau &= t , \\
    s_0(s,t) &= s e^{-t} , \\
    K(s,t) &= \mu s + \frac{1}{2} \sigma^2 s^2 + R(se^{-t}) 
    = K_\infty(s) + R(se^{-t}).
\end{align*}
As time tends to infinity the initial "excess" cumulants are lost:
\begin{align*}
\lim\limits_{t\to\infty}R(se^{-t})=R(0)=0 .
\end{align*}

We are now in a position to determine the time-evolution of the individual cumulants by differentiating with respect to $s$:
\begin{align*}
    \partial_sK(s,t)&=\mu+\sigma^2s+R'(se^{-t})e^{-t} , \\
    \partial_s^2 K(s,t)&=\sigma^2 + R''(se^{-t})e^{-2t} , \\
    \partial_s^{(k)}K(s,t) &=R^{(k)}(se^{-t})e^{-kt}, \quad k>2 .
\end{align*}
By evaluating at $s=0$, we have
\begin{align*}
    K_1(t) &= \mu + R_{1} e^{-t} , \\
    K_2(t) &= \sigma^2 + R_{2} e^{-2t} , \\
    K_k(t) &= R_{k}e^{-kt}, \quad k>2 ,
\end{align*}
where $R_1=\partial_s K(0,0)-\mu$, $R_2=\partial_s^2K(0,0) -\sigma^2$ and  $R_{k}=\partial_s^{(k)}K(0,0)=R^{(k)}(0)$ for $k>2$ are the initial residual cumulants. This shows that the decay rate of the $k^{th}$ cumulant is proportional to $k$, namely that the decay rate scales linearly with the order of the cumulant.
\end{enumerate}

\begin{figure}
    \centering
    \includegraphics[width=0.7\textwidth]{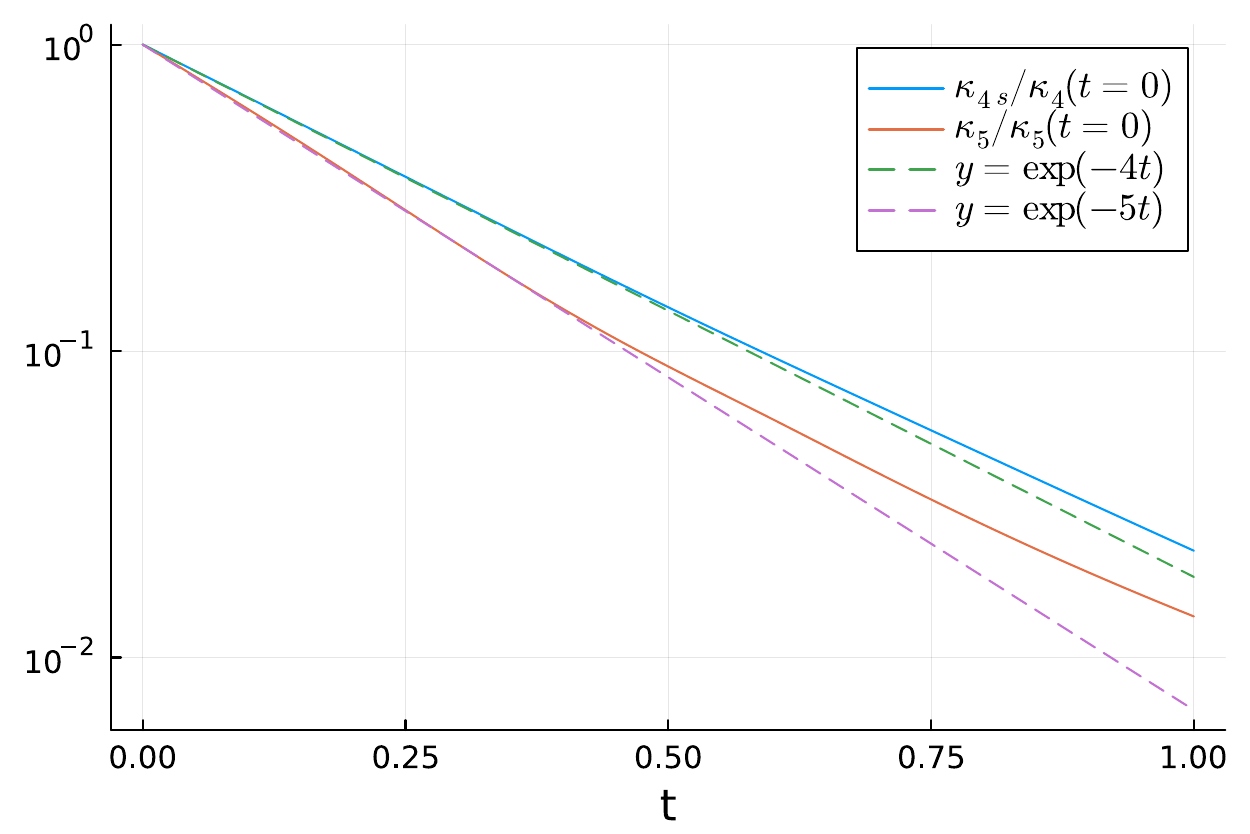}
    \caption{Decay of the fourth- and fifth-order cumulants, $\kappa_4$ and $\kappa_5$, over the course of a simulation where the initial distribution is taken as a sample of $N=1000$ particles from a double Maxwellian distribution. The cumulants are normalised to their initial value and analytic scaling laws are shown as dashed lines.}
    \label{fig:analytic_cumulant_decay}
\end{figure}

We check this scaling numerically in our method by fixing the $A_1$ and $A_2$ coefficients in equation \eqref{eq:velocity_ode}, instead of solving the system of equations shown in equations \eqref{eq:coefficient_lin_system}. Initialising the simulation with a double Maxwellian distribution, as shown in the second example, for $N=1000$ particles, $M = 41$ splines of order $4$, and a timestep of $\Delta t = 5 \times 10^{-3}$, we fix the coefficients to be $A_0 = 0$ and $A_1 = 1$. The evolution of the higher order cumulants is shown in Figure \ref{fig:analytic_cumulant_decay}. We observe that the cumulants scale at the predicted rate, until they reach the level of accuracy supported by the chosen resolution (approximately $10^{-2}$ for a particle resolution of $N=1000$. The saturation point of the cumulants corresponds to the solution reaching equilibrium. 


\bibliographystyle{jpp}

\bibliography{refs}

\end{document}